\documentstyle{mn}
\input{psfig.sty}
\title{Freely precessing neutron stars: Model and observations}

\author[D. I. Jones and N. Andersson] 
{D. I. Jones$^{1,2}$ and N. Andersson$^{1}$\\ 
$^{1}$ Faculty of Mathematical Studies, University of Southampton, 
       Highfield, Southampton, United Kingdom \\ 
$^{2}$ Department of Physics and Astronomy, University of Wales, 
       College of Cardiff, P.O.Box 913, Cardiff, United Kingdom \\ 
}

\begin{document}

\maketitle

\begin{abstract}

We present a model of a freely precessing neutron star which is then
compared against pulsar observations.  The aim is to draw conclusions
regarding the structure of the star, and test theoretical ideas of
crust-core coupling and superfluidity.  We argue that, on theoretical
grounds, it is likely that the core neutron superfluid does not participate
in the free precession of the crust.  We apply our model to the handful of
proposed observations of free precession that have appeared in the
literature.  Assuming crust-only precession, we find that all but one of
the observations are consistent with there not being any pinned crustal
superfluid at all; the maximum amount of pinned superfluid consistent with
the observations is about $10^{-10}$ of the total stellar moment of
inertia.  However, the observations do not rule out the possibility that
the crust and neutron superfluid core precess as a single unit.  In this
case the maximum amount of pinned superfluid consistent with the
observations is about $10^{-8}$ of the total stellar moment of inertia.
Both of these values are many orders of magnitude less than the $10^{-2}$
value predicted by many theories of pulsar glitches.  We conclude that
superfluid pinning, at least as it affects free precession, needs to be
reconsidered.

\end{abstract}


\section{Introduction}

This purpose of this study is to collect the important factors that
influence neutron star free precession, combine them into a single unified
model, and then test this model against observations.  In particular, we
wish to learn about the geological history of the star, decide whether or
not superfluid pinning in the inner crust is required to fit the data, and
ask whether it is the just the crust that participates in the free
precession, or whether some (or all) of the interior neutron superfluid
participates too.  By `observations' we refer to the handful of pulsars
where a smooth modulation in the pulse timing and/or structure has been
detected.  We hope that the model and equations presented here will be of
use to observers when new free precession candidates are discovered.

Clearly, we can only hope to extract useful information from the observations
if our model of neutron star free precession is sufficiently realistic.  To
this end, our model includes the effects of crustal elasticity, inertial
coupling (where the interior fluid pushes on the surrounding crust), and a
superfluid component pinned to the inner crust.  We also examine
dissipative processes that tend to enforce corotation between the crust
and core.  We will argue that current ideas concerning crust-core coupling
suggest that only the crust (and perhaps the charged plasma in the fluid
core) undergoes free precession, with the interior neutron fluid simply
rotating about an axis fixed in space.  This has been assumed implicitly in
earlier works \cite{as88}, but not commented on explicitly. 

The structure of this paper is as follows.  In section \ref{sect:dons} we
describe how strains in a neutron star crust can provide a non-spherical
contribution to the moment of inertia tensor.  In section \ref{sect:mfp} we
describe our free precession model.  Dissipative processes which convert
the precessional energy into heat or radiation are considered in section
\ref{sect:dm}.  In section \ref{sect:cfamwa} we describe how the finite
crustal breaking strain places limits on the shape and wobble angle of a
freely precessing star.  The way in which the signal from a pulsar is
modulated by free precession is described in section \ref{sect:eofpops}.
In section \ref{sect:aooofp} we combine the results of the previous
sections to extract as much information as possible from the proposed
observations of free precession that have appeared in the literature.  Our
conclusions are presented in section \ref{sect:conc}.

\section{Deformations of neutron stars}
\label{sect:dons}

To undergo free precession, a star must be deformed in some way, so that
its shape and moment of inertia tensor differ from that of an unstressed
fluid body.  In this section we describe how strain in the solid crust can
provide such a deformation, using the model of Pines \& Shaham (1972), who
in turn made use of the terrestrial analysis as presented in Munk \&
McDonald (1960).  For reasons of tractability a rotating but non-precessing
star is considered.  We will not consider deformations due to magnetic
stresses, although these could be important for the free precession of very
slowly rotating/very highly magnetised stars (see e.g. Melatos 1999).
Note, however, that a pulsar whose deformation was due entirely to magnetic
stresses would not display any modulation in its pulsations---see section
\ref{sect:tgc}.

Begin by writing down the total energy of the rotating star.  Let $I_{\rm
star}$ denote the moment of inertia of the spherical star, i.e. the moment
of inertia the star would have if it was unstrained and not rotating.  When
rotating, the moment of inertia about the rotation axis must be greater
than this, and so can be written as $I_{\rm star} (1+\epsilon)$.  We will
refer to $\epsilon$ as the \emph{total oblateness}.  We imagine the crust
to have solidified from a hot, liquid, state in the geologically distant
past, leaving it with a `reference' or zero-strain oblateness
$\epsilon_{0}$.  This parameter will then change only via crust-cracking or
a gradual plastic creep.  The star's energy is a function of $\epsilon$ and
$\epsilon_{0}$ according to:
\begin{equation} 
E = E_{\rm star} + \frac{J^{2} }{ 2I_{\rm star} (1+\epsilon) } 
          + A \epsilon^{2} + B(\epsilon - \epsilon_{0})^{2}.
\label{totalenergy}
\end{equation}
Here $E_{\rm star}$ denotes the energy the star would have if it was
spherical.  The second term is the kinetic energy.  The third term is the
increase in gravitational potential energy due to the star's shape no
longer being spherical.  The fourth is the elastic strain energy, which
depends quadratically on the difference between $\epsilon$, the actual
shape of the star, and $\epsilon_{0}$.  The constant $A$ depends on the
stellar equation of state, and will be of the order of the gravitational
binding energy of the star.  The constant $B$ also depends on the equation
of state, and will be of order of the total electrostatic binding energy of
the ionic crustal lattice.  The equilibrium configuration can be found by
minimising the energy at fixed angular momentum:
\begin{equation}
\label{onebulgeminim}
\left. \frac{\partial E}{\partial \epsilon}\right|_{J} = 0.
\end{equation}
For an entirely fluid star we would put $B=0$, giving an oblateness of
order of the ratio of kinetic and gravitational energies per unit mass:
\begin{equation}
\epsilon_{\rm fluid} \approx \frac{I_{\rm star} \Omega^2}{4A}.
\end{equation}
Given that $A$ is of the order of the gravitational binding energy we can
write this as:
\begin{equation}
\label{epsilonfluid}
\epsilon_{\rm fluid} \approx \frac{\Omega^{2} R^{3}}{GM}
      \approx 2.1 \times 10^{-3} \left(\frac{f}{100 \, \rm Hz}\right)^{2}
      R_6^3 / M_{1.4}
\end{equation}
where $\Omega =  2 \pi f$ is the angular frequency,  $R_6$ the neutron star
radius  in units  of $10^6$cm,  and  $M_{1.4}$ the  mass in  units of  $1.4
M_\odot$.

When the strain term is included we find
\begin{equation} 
\label{shape}
\epsilon =   \frac{I_{\rm star} \Omega^{2}}{4(A+B)} 
           + \frac{B}{A+B}\epsilon_{0}
      \equiv \epsilon_{\Omega} + b\epsilon_{0}.  
\end{equation} 
The oblateness is made up of two parts.  The first, $\epsilon_{\Omega}$,
scales as $\Omega^{2}$ and is due to centrifugal forces.  We will refer to
this as the \emph{centrifugal deformation}. The second term,
$b\epsilon_{0}$, is due entirely to the stresses of the crystalline solid,
and will be referred to as the \emph{Coulomb deformation}.  We have defined
$b = B/(A+B)$, which we will refer to as the \emph{rigidity parameter}.  It
is equal to zero for a fluid star ($B=0$) and unity for a perfectly rigid
one ($B/A \rightarrow \infty$).  Realistic neutron star equations of state
imply that $b$ takes a value of:
\begin{equation}
\label{bestimate}
b \approx 1.6 \times 10^{-5} R_6^5 / M_{1.4}^3.
\end{equation}
(See Jones (2000) for a simple derivation, and Ushomirsky, Cutler \&
Bildsten (2000) for a detailed numerical treatment).  Because this is
small, $b$ is approximately equal to $B/A$, and so is simply the ratio of
the crustal electrostatic binding energy to the total stellar gravitational
binding energy.  It is this second deformation that makes free precession
possible, not the (possibly much larger) centrifugal deformation.  We will
write the total change in the moment of inertia tensor due to this Coulomb
term as $\Delta I_{\rm d}$, so that
\begin{equation}
\label{DId}
\frac{\Delta I_{\rm d}}{I_{\rm star}} = b \epsilon_0.
\end{equation} 

Thus we see that the effect of elastic stresses in the crust is to change
the shape only slightly from that of the corresponding fluid body.
Physically, the smallness of this distortion is due to the Coulomb forces
being much smaller than the gravitational and centrifugal ones. 

As $b$ is small we have $\epsilon \approx \epsilon_{\Omega} \approx
\epsilon_{\rm fluid}$, so that in the following sections we will use
equation (\ref{epsilonfluid}) to estimate a star's actual oblateness.
Also, $\epsilon$ and $\epsilon_{0}$ can differ at most by the breaking
strain $u_{\rm break}$ of the crust, so we expect:
\begin{equation}
|\epsilon - \epsilon_{0}| \le u_{\rm break}.
\end{equation}
This breaking strain is very poorly constrained.  Estimates have ranged
from $\sim 10^{-2}$ to values very much lower---see section
\ref{sect:cfamwa}.

In the following section it will be useful to have an expression for the
ratio of the crustal to total stellar moments of inertia.  From Ravenhall
\& Pethick (1994) we find:
\begin{equation}
\label{Iratio}
\frac{I_{\rm crust}}{I_{\rm star}} 
\approx 1.5 \times 10^{-2} R_6^4 / M_{1.4}^2.
\end{equation} 
We will approximate the total stellar moment of inertia using the constant
density result
\begin{equation}
\label{Istar}
I_{\rm star} = \frac{2}{5} M R^2 = 1.12 \times 10^{45} 
               {\, \rm g \, cm^2\,} M_{1.4} R_6^2.
\end{equation}

\section{Modelling free precession}
\label{sect:mfp}

Real neutron stars will consist of an inelastic crust containing, in a
non-spherical cavity, a compressible liquid core.  This core will be made
up of a viscous electron-proton plasma, a neutron superfluid, and possibly
more exotic phases of matter also \cite{glen97}.  The crust itself may
contain a pinned superfluid in its inner parts.  Also, a magnetic field
will thread the star, in a way which depends on the properties of the
superfluid phase.  The superfluid core will couple in a frictional way to
the crust.  The free precession of such a system will clearly be far more
complicated than that of a rigid body.  The strategy that has been employed
to explore this free precession is to look at only one complicating factor
at a time.  Following this approach, we will briefly describe each
complicating feature.  In this section the effects of elasticity,
crust-core coupling and superfluid pinning are described, and then combined
to give a realistic free precession model.  All of these complications
preserve the kinetic energy of a precessing body.  The dissipative effects
of a frictional crust-core coupling and gravitational radiation reaction
are described in section \ref{sect:dm}.

\subsection{Rigid shell}
\label{sect:freeprec_rb}

\begin{figure}
\begin{picture}(100,100)(-50,0)

\thicklines

\put(0,0){\vector(0,1){80}}

\put(0,0){\vector(-2,3){40}}

\put(0,83){$\bf{J}$}

\put(0,0){\vector(1,4){13}}

\thinlines

\qbezier(-5.547,8.321)(-2.8,10.6)(0,10)


\qbezier(0,20.4)(2.5,21)(4.851,19.403)

\put(13,55){$\bf{\Omega}$}

\put(-40,63){$\bf{n_{d}}$}

\put(-3.5,11){$\bf{\theta}$}


\put(2.5,22){$\hat \theta$}

\end{picture}
\caption{This figure shows the reference plane for a freely precessing body,
which contains the deformation axis $ {\bf n_{d}}$, the angular velocity
vector ${\bf \Omega}$ and the fixed angular momentum ${\bf J}$.  The
vectors ${\bf n_{d}}$ and ${\bf \Omega}$ rotate around $\bf{J}$ at the
\emph{inertial precession frequency} $\dot{\phi}$.  We refer to $\theta$ as
the \emph{wobble angle}.}
\label{refplane}
\end{figure}
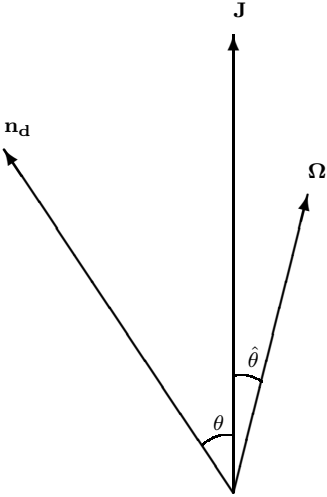

We will begin by describing the simple case of a rigid shell.  The moment
of inertia tensor of any axisymmetric rigid body can be written as
\begin{equation}
{\bf I} =    I_0 \bdelta + \Delta I_{\rm d} ({\bf n_d n_d} 
         - \bdelta/3),
\end{equation}
where the unit vector ${\bf n_d}$ points along the body's symmetry axis. 
Then the principal moments are $I_{1} = I_{2} = I_0 -\Delta I_{\rm
d}/3$, $I_{3} = I_0 + 2\Delta I_{\rm d}/3$, so that $I_{3}-I_{1} =
\Delta I_{\rm d}$.  The angular momentum is then related to the angular
velocity according to
\begin{equation}
{\bf J} = (I_0 - \Delta I_{\rm d}/3) 
          {\bf \Omega} - \Delta I_{\rm d} \Omega_{3} {\bf n_d},
\end{equation}
where the 3-axis lies along ${\bf n_d}$.  This shows that the three vectors
${\bf J, \Omega}$ and ${\bf n_d}$ are always coplanar.  Following Pines \&
Shaham (1972) we will call the plane so defined the \emph{reference plane}
(see figure \ref{refplane}).
Given that the angular momentum is fixed, this plane must revolve around
${\bf J}$.  The free precession is conveniently parameterised by the angle
$\theta$ between ${\bf n_{\rm d}}$ and {\bf J}.  We will refer to this as
the \emph{wobble angle}.  For a nearly spherical body the angle
$\hat{\theta}$ between ${\bf \Omega}$ and ${\bf J}$ is much smaller than
the angle between ${\bf J}$ and ${\bf n_d}$, according to
\begin{equation}
\label{eq:rigid_theta_hat}
\hat{\theta} \approx \frac{\Delta I_{\rm d}}{I_{1}} \sin \theta \cos \theta.
\end{equation}
We will denote by ${\bf n_{J}}$ the unit vector along
${\bf J}$.  Decomposing the angular velocity according to
\begin{equation}
{\bf \Omega} = \dot{\phi} {\bf n_{J}} + \dot{\psi} {\bf n_d}
\end{equation} 
then gives 
\begin{equation}
J = I_{1} \dot{\phi},
\end{equation}
\begin{equation}
\label{psidotrigid}
\dot{\psi} = - \frac{\Delta I_{\rm d}}{I_{3}} \dot{\phi}.
\end{equation}
The symmetry axis ${\bf n_d}$ performs s a rotation about ${\bf J}$ in a
cone of half-angle $\theta$ at the angular frequency $\dot{\phi}$.  We will
refer to this as the \emph{inertial precession frequency}.  There is a
superimposed rotation about the symmetry axis ${\bf n_d}$ at the angular
velocity $\dot{\psi}$ .  This is usually referred to as the \emph{body
frame precessional frequency}, with the corresponding periodicity known as
the \emph{free precession period}:
\begin{equation}
P_{\rm fp} = \frac{2\pi}{\dot \psi}
\end{equation}
For a nearly spherical body equation (\ref{psidotrigid}) shows that
$\dot{\psi} \ll \dot{\phi}$, or equivalently $P \ll P_{\rm fp}$.  Note that
the angles $(\theta, \phi, \psi)$ are simply the usual Euler angles which
describe the orientation of the rigid body (see e.g. Landau \& Lifshitz
1976, figure 47).

In subsequent sections we will find it necessary to make the approximations
of small wobble angle and nearly spherical stars.  In this case equations
(\ref{eq:rigid_theta_hat}) and (\ref{psidotrigid}) become:
\begin{equation}
\label{approxthetahat}
\hat{\theta} \approx \frac{\Delta I_{\rm d}}{I_0} \theta,
\end{equation}
\begin{equation}
\label{approxpsidot}
\dot{\psi} \approx - \frac{\Delta I_{\rm d}}{I_0} \dot{\phi}.
\end{equation}

\subsection{Elastic shell}
\label{sect:efp}

Following section \ref{sect:dons}, we will write the moment of inertia
tensor of a rotating elastic shell as the sum of a spherical and two
quadrupolar parts:
\begin{equation}
\label{eq:elasticmoi}
{\bf I} = I_0 \bdelta 
         + \Delta I_{d} ({\bf n_{d}n_{d}} - \bdelta/3)
         + \Delta I_{\Omega} ({\bf n_{\Omega}n_{\Omega}} - \bdelta/3).
\end{equation}
(See Pines \& Shaham (1972) and Munk \& McDonald (1960) for further
discussion).  The first term on the right hand side is the moment of
inertia of the non-rotating undeformed spherical shell.  The second term is
the change due to crustal Coulomb forces, and has the unit vector ${\bf
n_{d}}$, fixed in the crust, as its symmetry axis.  The third term is the
change due to centrifugal forces, and has ${\bf n_{\Omega}}$, the unit
vector along ${\bf \Omega}$, as its symmetry axis.  When $\bf \Omega$ moves
with respect to the body the shell changes shape.  This is why the shell is
described as elastic.

When the directions ${\bf n_{d}}$ and ${\bf n_{\Omega}}$ coincide the
body spins about its symmetry axis without precessing.  When the two
directions differ the body will precess with triaxial shape.  
Proceeding exactly as in the rigid body case, it is easy to prove that
the elastic body undergoes a free precessional motion like that of a
rigid \emph{axisymmetric} body, with what we will call an
\emph{effective moment of inertia tensor} given by:
\begin{equation}
I_{1} = I_0 - \Delta I_{d}/3 + 2\Delta I_{\Omega}/3,
\end{equation}
\begin{equation}
I_{2} = I_{1},
\end{equation}
\begin{equation}
I_{3} = I_0 + 2\Delta I_{d}/3 + 2\Delta I_{\Omega}/3.
\end{equation}
Crucially, the centrifugal piece of the quadrupole moment enters in a
spherical way---it is only the $\Delta I_{\rm d}$ piece that is responsible
for the free precession.  In particular, equations
(\ref{eq:rigid_theta_hat}) and (\ref{psidotrigid}) still apply, with $I_1$
and $I_3$ as given above, and with the precession frequency and angle $\hat
\theta$ both proportional to the Coulomb deformation $\Delta I_{\rm d}$.

\subsection{Rigid shell containing non-spherical fluid}
\label{sect:rscnf}

First consider the free precession of a rigid axisymmetric shell containing
a \emph{spherical} fluid cavity.  We will assume that there are no viscous
or frictional interactions between the shell and the fluid.  In this case
the free precession of the shell is exactly the same as if the fluid were
not there.  However, if the cavity is non-spherical, there will be a
reaction force between the rigid shell and the fluid, due to the fluid
tending toward a configuration symmetric about its rotational axis, and
therefore `pushing' on the shell.  This pushing is known as \emph{inertial
coupling}.  In the case of a homogeneous incompressible fluid, the combined
motion of fluid and shell is given in Lamb's monograph (1952).  A number of
simplifying assumptions are necessary: The motion of the fluid is always
one of uniform vorticity; the ellipticity of the shell and cavity are
small; and the wobble angle $\theta$ is small.

It is then found that the shell undergoes the usual free precession motion,
so that equations (\ref{approxthetahat}) and (\ref{approxpsidot}) apply,
with $\Delta I_{\rm d}$ equal to the difference between the 1 and 3
principal moments of inertia \emph{of the whole star}, not just the shell,
while $I_0$ refers to the shell only.  The system's total angular momentum
${\bf J}$ is simply the sum of the angular momenta of the shell and fluid,
both of which remain very nearly parallel to ${\bf J}$.  As defined
previously, $\hat{\theta}$ is the angle that the angular velocity of the
shell makes with ${\bf J}$.  Crucially, the angular velocity of the fluid
remains very close to the fixed total angular momentum of the system,
i.e. \emph{inertial coupling does not cause the fluid to participate in the
free precessional motion of the shell}.

\subsection{Rigid shell with pinned superfluid}

According to many neutron star models which attempt to explain
post-glitch behaviour, the neutron vortices which coexist with the
inner crust become pinned, at many points along their length, to
nuclei in the crustal lattice.  The velocity field of this pinned
superfluid is specified entirely by the distribution of these pinning
sites.  The pinning sites themselves are rigidly fixed to the crust.
It follows that if a rigid crust is set into free precession, the
instantaneous velocity field of the pinned superfluid continually
adjusts, according to the orientation of the crust.  Such a
precessing star was studied by Shaham (1977).  If we  write down
the angular momentum of the crust, including that of its pinned
superfluid, we have
\begin{equation}
\label{eq:jsf}
{\bf J} = {\bf I \Omega} + {\bf J_{SF}}.
\end{equation}
The quantities ${\bf I}$ and ${\bf \Omega}$ refer to the crust only, while
${\bf J_{SF}}$ is the angular momentum of the pinned superfluid.  The
orientation of ${\bf J_{SF}}$ is fixed with respect to the crust.  We
will consider only the simplest case, where all the pinned superfluid
points along the crust's deformation axis, so that ${\bf J_{SF}} =
J_{\rm SF} {\bf n_d}$.  Then repeating the analysis of section
\ref{sect:freeprec_rb} we find (for nearly spherical bodies with $\theta
\ll 1$)
\begin{equation}
\label{psidotsf}
\dot{\psi} = - \frac{\Delta I}{I_0} \dot{\phi} 
             - \frac{J_{\rm SF}}{I_0},
\end{equation}
and
\begin{equation}
\hat{\theta} \approx \left[   \frac{\Delta I}{I_0} 
                            + \frac{J_{\rm SF}}{J_{3}} \right] \theta.
\end{equation}
In words, the pinned superfluid acts so as to increase the effective
non-sphericity of an oblate body, increasing the magnitude of both the
body frame precession frequency, and the misalignment between the
crust's angular momentum and angular velocity vectors.

The rotation rate of the pinned superfluid will probably be very close to
the star's rotation rate, as even a small difference between the two would
give rise to a Magnus force which would break the pinning \cite{lg98}.  We
can therefore put $J_{\rm SF} \approx I_{\rm SF} \dot{\phi}$, where $I_{\rm
SF}$ is the moment of inertia of the pinned superfluid.  Then the above
equations can be written very simply as:
\begin{equation}
\label{eq:psi_dot_eff}
\dot{\psi} = - \left[\frac{\Delta I_{\rm d}}{I_{0}} 
                     + \frac{I_{\rm SF}}{I_{0}} \right]
               \dot{\phi}
\end{equation}
\begin{equation}
\label{eq:theta_hat_eff}
\hat{\theta}  \approx \left[ \frac{\Delta I_{\rm d}}{I_{0}} 
            + \frac{I_{\rm SF}}{I_{0}} \right] \theta.
\end{equation}
We will call the term in square brackets the \emph{effective oblateness
parameter}.  It is made up of both crustal distortion and pinned superfluid
parts.

\subsection{Composite model}
\label{sect:cm}

We now wish to consider the free precession of a more realistic composite
model---an elastic shell containing a fluid cavity, with superfluid pinned
to the shell.  We would then form the equation of motion of the body by
combining equations (\ref{eq:elasticmoi}) and (\ref{eq:jsf}).  When
inertial coupling forces are neglected it is straightforward to repeat the
analysis to show that the effects of elastic deformation and superfluidity
add in a simple way.  (Equations
(\ref{eq:psi_dot_eff})--(\ref{eq:theta_hat_eff}) apply, with $\Delta I_{\rm
d}$ the Coulomb deformation of the shell and $I_0$ the crustal moment of
inertia).  The effect of also including inertial coupling forces will be to
set the $\Delta I_{\rm d}$ factor to the deformation in the moment of
inertia of the \emph{whole star}, not just the change in the crustal moment
of inertia.  Then equations of the form
(\ref{eq:psi_dot_eff})--(\ref{eq:theta_hat_eff}) apply again.  In full:
\begin{equation}
\label{comppsidot}
\dot \psi = - \epsilon_{\rm eff} \dot \phi,
\end{equation}
\begin{equation}
\label{compthetahat}
\hat \theta  = \epsilon_{\rm eff} \theta,
\end{equation}
\begin{equation}
\label{compeeff}
\epsilon_{\rm eff} = \frac{\Delta I_{\rm d}}{I_{0}} + \frac{I_{\rm SF}}{I_{0}},
\end{equation}
where $I_{0}$ is equal to the crustal moment of inertia only, while $\Delta
I_{\rm d}$ is the Coulomb-induced deformation in the moment of inertia of
the whole star, and $I_{\rm SF}$ is the moment of inertia of the pinned
superfluid.

\section{Dissipation mechanisms}
\label{sect:dm}

The model described above would, once excited, precess forever, as no
dissipative energy losses have been included.  A real star will suffer a
number of such losses.  To complete our model we will therefore consider
two types of dissipation: A frictional crust-core coupling, and
gravitational radiation reaction.

Before considering these particular cases we will derive a general
expression for the damping timescale. The energy of the precessing
crust can always be written as a function of its angular momentum and
wobble angle.  For wobble damping the angular momentum of the crust is
nearly constant, so we can write:
\begin{equation}
\dot{\theta} = \dot{E}  {\left/ \frac{\partial E}{\partial \theta}\right|_J}.
\end{equation}
Here $E$ denotes the total energy of the shell plus pinned superfluid,
and will include kinetic, elastic and gravitational parts.  Cutler \&
Jones (2000) have shown that, to leading order, only the kinetic energy
need be considered.  It is then straightforward to take the free
precession model of section \ref{sect:cm} and evaluate the partial
derivative of the denominator, giving
\begin{equation}
\dot{\theta} = \frac{\dot E}{\dot{\phi}^{2}\theta I_{0} \epsilon_{\rm eff}}.
\end{equation}
Expressed as a damping time this is
\begin{equation}
\tau_{\theta} = -\frac{\theta}{\dot \theta}
              = -\frac{\dot{\phi}^{2}\theta^{2}I_{0} 
                 \epsilon_{\rm eff}}{\dot E}.
\end{equation}

The quantity $\dot{E}$ will always be negative, corresponding to the
conversion of mechanical energy into heat or radiation.  It therefore
follows that dissipation tends to decrease the wobble angle of stars with
oblate deformations, but tends to increase the wobble angle of stars with
prolate deformations \cite{cj00}.  The crustal strains considered in this
paper will almost certainly lead only to oblate deformations.  More exotic
scenarios (perhaps a very strong \emph{toroidal} magnetic field) might lead
to prolate ones.  Dissipation in such stars would eventually lead to the
deformation axis being orthogonal to the spin axis.  Such a non-precessing
triaxial star would then spin down gravitationally.

\subsection{Frictional crust-core coupling}

We have already considered one crust-core interaction, namely inertial
coupling, due to the core fluid simply `pushing' on the precessing shell.
However, in a real star there will be additional crust-core interactions,
which have been investigated by theorists attempting to explain the
post-glitch behaviour of pulsars.  The core itself consists of two
coexisting fluids: a plasma of electrons and protons, and a neutron
superfluid.  The plasma makes up only a few percent of the total mass.

The crust-core interaction proceeds in two stages.  In the first stage the
crust couples to the core plasma.  This coupling is mediated by two
separate interactions: plasma viscosity, and an electromagnetic coupling,
where Alfv\'{e}n waves communicate the crust's motion to the interior.  In
the second stage, the plasma couples to the neutron superfluid, due to the
scattering of electrons off the superfluid vortices. (Strictly, the
electrons are scattered by the magnetic field created by protons which are
entrained around the vortex cores; see Alpar \& Sauls 1988).  This second
interaction is frictional in nature, i.e. is a drag force proportional to
the velocity difference between the electrons and the vortices.  It is
sometimes referred to as `mutual friction'.

There will be energy losses associated with both couplings.  However,
following the work of Easson (1979) it is usually assumed that the
crust-plasma coupling timescale is much \emph{less} than the frictional
one, so that the crust and core plasma can be treated as a single system,
interacting frictionally with the neutron core.  We will work in this
limit.  (Relaxation of this assumption will in fact tend to strengthen our
conclusion, namely that the neutron superfluid does not follow the
precession of the crust).

This frictional interaction is important in two regards.  Firstly, it will
lead to a dissipation of the precessional energy.  Secondly, if strong
enough, it would cause the \emph{whole} star---crust \emph{and} core---to
precess as a single body.  We can therefore imagine two extreme cases.
When the frictional coupling is very weak the crust precesses on top of a
non-precessing core, with only the inertial forces of section
\ref{sect:rscnf} coupling the two.  Then equation (\ref{compeeff}) applies,
with $I_0 = I_{\rm crust}$.  At the other extreme, when the frictional
forces are very strong, the star precesses as a single unit. Then equation
(\ref{compeeff}) again applies, but with $I_0 = I_{\rm star}$.  It is
clearly important to know where on this scale real neutron stars can be
expected to be found, both from the point of view of modelling the
precession, and for estimating the rate at which it is damped.

The motion of a rigid shell containing a spherical cavity, with a
frictional coupling acting between the two, was invested by Bondi \& Gold
(1955).  They made the simplifying assumption that the motion of the fluid
was one of rigid rotation.  The shell is acted upon by a torque
\begin{equation}
{\bf T} = K({\bf \Omega_{fluid} - \Omega_{solid}}),
\end{equation}
where $K$ is a positive constant.  An equal and opposite torque acts on the
fluid. The equations of motion are then
\begin{equation}
\label{eq:bg}
\frac{d \bf J_{shell}}{dt} 
   = K({\bf \Omega_{fluid} - \Omega_{solid}})
   = -\frac{d \bf J_{fluid}}{dt}
\end{equation}
Bondi \& Gold found the normal modes of this two component system for small
wobble angles.  The results below can be readily found from their solution.

First consider the simple case of the non-precessing shell and fluid,
rotating about the same axis, but at different rates.  Using the component
of the above equation along the shell's symmetry axis, it is easy to show
that the relative rotation rate decreases exponentially.  We will write the
e-folding time of this decay as $n$ rotation periods, corresponding to a
time $2\pi n/ \Omega$. 

Using the components of (\ref{eq:bg}) orthogonal to the symmetry axis it
can then be shown that in the case $n \ll 1$ the fluid is tightly coupled
to the shell, and such a body would rotate and precess as if it were a
single uniform solid.  In the case $n \gg 1$ the fluid is only loosely
coupled, so that while the shell undergoes free precession, the fluid's
angular velocity vector remains very nearly fixed in space.

Alpar, Langar \& Sauls (1984) calculated the coupling strength for the
electron-vortex core interaction described above.  Using this result, Alpar
\& Sauls (1988) estimated $n$ to lie in the interval $400 \rightarrow
10^{4}$.  As $n \gg 1$ it follows at once that real neutron stars lie in
the weak coupling regime.  In this regime it can then be shown (using the
components of (\ref{eq:bg}) orthogonal to the symmetry axis) that if the
shell is set into free precession, with its angular momentum vector
remaining along the spin axis of the fluid, the free precession is damped,
with an e-folding time of $n$ body frame free precession periods, i.e. in a
time $2 \pi n/ \dot{\psi}$.

In summary, \emph{this frictional interaction is too weak to cause the core
neutron superfluid to participate in the free precession of the star's
crust.  Instead, it serves only to damp the free precession of the crust,
on a timescale of between $400$ and $10^4$ free precession periods.}

\subsection{Gravitational radiation reaction}

The effect of gravitational radiation reaction on precessing elastic bodies
with a fluid non-spherical core was recently considered by Cutler \& Jones
(2000).  It was found that the wobble angle $\theta$ was damped
exponentially on a timescale
\begin{equation}
\tau_{\rm g} = \frac{5c^5}{2G} \frac{I_0}{(\Delta I_{\rm d})^2}
               \frac{1}{\Omega^4}.
\end{equation}
This remains valid when the effects of superfluid pinning are included.
Parameterising:
\begin{equation}
\tau_{\rm g} = 1.8 \times 10^{6} \, {\rm yr \,}
\left(\frac{10^{43}{\rm g \, cm^{2}}}{I_0}\right)
\left(\frac{10^{-6}}{\Delta I_{\rm d}/I_0}\right)^{2} 
\left(\frac{P}{1 {\rm ms}}\right)^{4}.
\end{equation}
For instance, for a star with $\Delta I_{\rm d}/I_0 = \epsilon_{\rm eff} =
10^{-6}$, $I_0 =10^{43}{\rm g \, cm^{2}}$ and $P=1$ms, this corresponds to
damping in $n= 6 \times 10^9$ free precession periods.  Comparing this with
the damping rate due to friction, we see that gravitational radiation
reaction is not an important source of wobble damping in any neutron star
of physical interest.  (This does not mean that the \emph{spin-down}
component of the radiation reaction is unimportant---see section
\ref{sect:aoo}).

\section{Crust fracture and maximum wobble angle}
\label{sect:cfamwa}

A real neutron star crust will have a finite breaking strain $u_{\rm
break}$.  The actual value of this breaking strain is highly uncertain.  As
discussed by Ruderman (1992), extrapolations from laboratory crystals
suggest that values as high as $10^{-2}$ are possible, but the actual value
may be much lower.  Ruderman suggests a value of $10^{-4}$ as plausible.
For an axisymmetric star this breaking strain will limit the difference
between its actual shape, which is approximately $\epsilon_{\rm fluid}$,
and the reference shape of the crust, $\epsilon_0$.  However, when a star
is set into free precession, additional time-dependent strains will result,
even if the star is initially `relaxed' (i.e. $\epsilon_{\rm fluid} =
\epsilon_0$).  In general, a precessing star's crust will suffer both sorts
of strain \cite{ps72}.

First consider a non-precessing spinning-down star.  Its crust presumably
solidified (and $\epsilon_0$ was fixed) when it was spinning more rapidly
that at present. This means its reference oblateness $\epsilon_0$ is
greater than its current actual oblateness $\epsilon_{\rm fluid}$, so that
we would now expect $\epsilon_{0} > \epsilon_{\rm fluid}$.  Combining with the
bound due to crust fracture we then have
\begin{equation}
\label{ineqobs}
0 < \epsilon_{0} - \epsilon_{\rm fluid} < u_{\rm break}.
\end{equation}
Even if the crust has cracked a number of times during spin-down we would
still expect the star to have some `memory' of its initial shape (unless
the cracking was able to relieve \emph{all} the stresses in the crust), so
that the above equation should still hold.

In the case of an accreting star the situation is different.  If the star
has been spun up from a relaxed state at a small rotation rate we would
expect its current oblateness $\epsilon_{\rm fluid}$ to \emph{exceed} its
reference oblateness $\epsilon_0$.  However, the accretion will tend to
create new crustal material relaxed to the current rotation rate of the
star, so that the difference $\epsilon_{\rm fluid} - \epsilon_{0}$ may be
rather small.

We will now look at the opposite case, when the strains are \emph{due to
precession only}, i.e. when $\epsilon_0 = \epsilon_{\rm fluid}$.  A simple
treatment is possible, based upon the known geometry of free precession for
an elastic body, as discussed in section \ref{sect:efp}.  These strains can
be used to place a limit on the maximum possible wobble angle $\theta$ a
star can sustain.

We know that (for small wobble angles at least) when an initially axially
symmetric body is set into free precession a deformation $\Delta I_{\rm d}$
remains along the axis ${\bf n_{d}}$ fixed in the star, while a deformation
$\Delta I_{\Omega}$ points along the angular velocity vector.  From the
point of view of an observer attached to the crust, a deformation of size
$\Delta I_{\Omega}$ describes a cone of half-angle $\theta + \hat{\theta}
\approx \theta$ about ${\bf n_d}$.  This change in shape is all we
need to know to estimate the strain: The change in position of any given
particle is of order $R \epsilon_{\Omega} \theta$, while the corresponding
strain is of order $\epsilon_{\Omega} \theta$.  This precession-induced
strain is not constant, but varies with magnitude $\epsilon_{\Omega}
\theta$ over one (body frame) free precession period.  As there exists a
maximum strain $u_{\rm break}$ that the solid can withstand prior to
fracture, the wobble angle will be limited to a value of $u_{\rm
break}/\epsilon_{\Omega} \approx u_{\rm break}/\epsilon_{\rm fluid}$ so
that:
\begin{equation}
\label{eq:thetamax}
\theta_{\rm max} \approx 0.45 \left(\frac{100 \, \rm Hz}{f}\right)^{2} 
                 \left(\frac{u_{\rm break}}{10^{-3}}\right) {\rm \, radians}.
\end{equation}
Qualitatively, we can say that faster spinning neutron stars have larger
bulges to re-orientate and therefore can sustain smaller wobble angles
prior to fracture.  For sufficiently slowly spinning stars the above
equation breaks down, yielding angles in excess of $\pi/2$.  The wobble
angles of such slowly spinning stars are not limited by crustal strain.  To
give two extreme examples, for a breaking strain of $10^{-3}$ the wobble
angle of a 300Hz neutron star in a low-mass X-ray binary or millisecond
pulsar would be limited to about $3^{\circ}$, while a `standard' field
pulsar spinning at around a Hz could precess with \emph{any} wobble angle.

We therefore have two separate bounds that a freely precessing star must
satisfy, one due to the mismatch between its reference and actual shape,
and one due to free precession.  These bounds will be of use in section
\ref{sect:aooofp}, when we examine possible observations of free
precession.

\section{Effect of free precession on the pulsar signal}
\label{sect:eofpops}

In this section we will examine the effect of free precession on the
electromagnetic signal of a pulsar.  The aim is to describe how free
precession might be detected using electromagnetic data, by examining
variations in the pulse frequency, amplitude and polarisation.  From these
we hope to test our free precession model and extract information
concerning the wobble angle, effective oblateness and superfluid pinning.

In section \ref{sect:efpps:ge} we will consider the problem of how the
precession affects the electromagnetic pulses for a torque-free top.  This
calculation requires only the geometry of free precession, and we will
refer to these modulations as the purely geometric modulations.  However,
real pulsars are acted on by electromagnetic spin-down torques.  The
magnitude of these torques will themselves be modulated by the free
precession, although the form of the modulation depends on the model of
spin-down torque employed.  These torques will, in turn, modify the
pulsations.  In section \ref{sect:efpps:et} we will include these variable
torques using a simple spin-down model.

Some of the results given here have been presented previously, in varying
degrees of generality.  We hope that by assembling all of the important
observational characteristics of free precession in one place, we will
provide a resource for observers wishing to asses the likelihood of a given 
pulse modulation being caused by free precession.

\subsection{Effect of free precession on the pulsar signal: Geometric effects}
\label{sect:efpps:ge}

We will begin by modelling the neutron star as a torque-free precessing
symmetric top.  We will model the pulsations in the obvious way, i.e. as a
thin beam fixed with respect to the star's body axes, aligned with the
dipole moment ${\bf m}$.  Each pulsation then corresponds to the passage of
${\bf m}$ through the plane containing the angular momentum vector ${\bf
J}$ and the observer.  The motion of ${\bf m}$ is then a slow rotation at
$\dot{\psi}$ about ${\bf n_{d}}$, the deformation axis, superimposed on the
rapid rotation of ${\bf n_{d}}$ at $\dot{\phi}$ about ${\bf J}$.  This will
lead to variations in the pulse phase, amplitude and polarisation, on the
body frame precession timescale $P_{\rm fp} = 2\pi/\dot\psi$.  We will
begin by considering the phase variations.

\subsubsection{Phase modulation}

The problem of phase modulation due to precession was first considered
by Ruderman (1970), and has been elaborated upon since
by \nocite{bms90} Bisnovatyi-Kogan et al. (1990) and Bisnovatyi-Kogan
\& Kahabka (1993) \nocite{bk93} in connection with the 35 day
periodicity observed in Her X-1.  Following the discovery of planets
around pulsar PSR 1257+12 a number of authors re-examined the issue to
check whether or not free precession could mimic planetary
perturbations of the pulsation \cite{nfw90,cord93,gjk93,glen95}. We
will comment upon their findings in section \ref{sect:efpps:et}.

Let ${\bf \hat{m}}$ denote a unit vector along ${\bf m}$.  Denote the
orientation of the body frame axes $\{\hat{x},\hat{y},\hat{z}\}$ with
respect to the inertial frame axes $\{x,y,z\}$ by the usual Euler angles
$(\theta,\phi,\psi)$.  (See figure 47 of Landau \& Lifshitz 1976).  Let
${\bf \hat{m}}$ lie in the $\hat{x} \hat{z}$ plane, at an angle $\chi <
\pi/2$ to the $\hat{z}$-axis.  Then the components $[\hat m_x, \hat m_y,
\hat m_x]$ of ${\bf \hat{m}}$, referred to the inertial frame are
\[
\label{mhat}
\left[
\begin{array}{c}  \cos \phi \cos \psi \sin \chi  
                - \sin \phi \cos \theta \sin \psi \sin \chi 
                + \sin \phi \sin \theta \cos \chi \\
                  \sin \phi \cos \psi \sin \chi 
                + \cos \phi \cos \theta \sin \psi \sin \chi 
                - \cos \phi \sin \theta \cos \chi \\
                  \sin \theta \sin \psi \sin \chi + \cos \theta \cos \chi
\end{array}
\right]
\]
Define Euler-like angles $\Theta$ and $\Phi$ to describe the polar angle
and azimuth of ${\bf \hat{m}}$.  Then the $\Phi$ angle describes the
phase of the pulsar signal, with a pulsation being observed whenever
$\Phi$ is equal to the azimuth of the observer, e.g. whenever $\Phi =
0$ for an observer in the inertial $x>0, z>0$ quarter-plane.  Then
\begin{equation}
\tan \Phi =  \frac{\hat{m}_{y}}{\hat{m}_{x}}.
\end{equation}
A little algebra leads to 
\begin{equation}
\label{Phi}
\Phi =  \phi - \frac{\pi}{2} + \arctan \left[ \frac{1}{\cos \theta}
        \left( \frac{\cos \psi \tan{\chi}}{\tan \theta - \sin \psi \tan \chi} 
        \right) \right]
\end{equation}
and also
\begin{equation}
\label{Psid}
\dot{\Phi}  =  \dot{\phi} 
\end{equation}
\[
\hspace{0.5cm} + \dot{\psi} \sin \chi \frac{\cos \theta \sin \chi 
           - \sin \psi \sin \theta \cos \chi}
           {(\sin \theta \cos \chi - \cos \theta \sin \psi \sin \chi)^{2} 
           + \cos^{2}\psi \sin^{2}\chi},
\]
where $\dot{\Phi}$ is the \emph{instantaneous electromagnetic frequency}.
Its time-averaged value is what we would normally refer to as the `spin
frequency' of the star, which we will denote by $\Omega$.  In order to
proceed it is necessary to treat the $\theta > \chi$ and $\theta < \chi$
cases separately.

\subsubsection*{\bf $\theta > \chi$ \lowercase{case}}
\label{sect:tgc}

First consider the \emph{average} electromagnetic pulse frequency.  An
increase of $\phi$ by $2 \pi$ at fixed $\psi$ causes ${\bf \hat{m}}$ to
rotate once about ${\bf J}$, i.e. causes a pulsation.  However, an increase
in $\psi$ by $2 \pi$ at fixed $\phi$ \emph{does not} rotate ${\bf \hat{m}}$
about ${\bf J}$, i.e. does not cause a pulsation.  It follows that the
\emph{average} electromagnetic frequency is exactly $\dot{\phi}$.  The
departure from this average spin rate can best be expressed as a phase
residual $\Delta \Phi$:
\begin{eqnarray}
\Delta \Phi & = & \Phi - (\phi - \frac{\pi}{2}) \\
            & = & \arctan \left[ \frac{1}{\cos \theta}
                  \left( \frac{\cos \psi \tan{\chi}}
                  {\tan \theta - \sin \psi \tan \chi}\right) \right].
\label{DParbitrary}
\end{eqnarray}
The denominator of the term in curved brackets is clearly non-zero for all
$\psi$, and so $\Delta \Phi$ remains in the range $-\pi/2 < \Delta \Phi <
\pi/2$ \cite{nfw90}.  In the case $\theta \gg \chi$ we find
\begin{equation}
\Delta \Phi = \frac{\chi}{\sin \theta} \cos \psi
\end{equation}
and
\begin{equation}
\Delta \dot\Phi = -\dot\psi \frac{\chi}{\sin \theta} \sin \psi
\end{equation}
When $\chi =0$, as would be the case for a star whose deformation is due
entirely to axisymmetric magnetic stresses, there is no modulation in the
pulsations at all.  The free precession of such a star would only be
detectable is there was some non-axisymmetry in the pulsar beam.

\subsubsection*{\bf $\theta < \chi$ case}

As described in section \ref{sect:cfamwa}, the wobble angles of rapidly
rotating stars are limited to small values by the finite crustal breaking
strain, so that for such stars this is almost certainly the case of
interest.  Again begin by considering the average pulsation frequency.  An
increase of $\phi$ by $2 \pi$ at fixed $\psi$ causes ${\bf \hat{m}}$ to
rotate once about ${\bf J}$, i.e. causes a pulsation.  It is also the case
that an increase of $\psi$ by $2 \pi$ at fixed $\phi$ causes ${\bf
\hat{m}}$ to rotate once about ${\bf J}$, i.e. causes a pulsation.  It
follows that the average pulsation frequency is $\dot{\phi} + \dot{\psi}$.
The phase residual is now given by
\begin{equation}
\label{generalDP}
\Delta \Phi = \Phi - (\phi + \psi) 
\end{equation}
\[ 
\vspace{0.5cm} =  \arctan \frac{(\cos \theta -1)\sin \psi \sin \chi 
                - \sin \theta \cos \chi}
                 {\cos \psi \sin \chi + (\cos \theta \cos \psi \sin \chi 
                - \sin \theta \cos \chi) \tan \psi}. 
\]
It is straightforward to show that the denominator of this function is
non-zero for all $\psi$, and so $\Delta \Phi$ remains in the range
$-\pi/2 < \Delta \Phi < \pi/2$ \cite{nfw90}.

When $\theta \ll \chi$ we find
\begin{equation}
\label{DPsmalltheta}
\Delta \Phi = - \theta \frac{\cos \chi}{\sin \chi} \cos \psi
\end{equation}
and 
\begin{equation}
\label{DPdsmalltheta}
\Delta \dot{\Phi} = \dot{\psi}  \theta \frac{\cos \chi}{\sin \chi} \sin \psi.
\end{equation}
We therefore see that for small wobble angles the phase residual varies
sinusoidally on the (long) free precession timescale, with an amplitude
$\theta$.  The fractional variation in pulsation frequency is of order
$\theta \dot{\psi}/\dot{\phi} \sim \theta \epsilon_{\rm eff}$.

\subsubsection{Amplitude modulation}

The precessional motion will modulate the amplitude of the pulsar signal,
although the precise form of the modulation will depend upon the geometry
of the emission region.  Assuming emission axisymmetry about the dipole
axis and that the intensity of emission falls off over an angular width
$W$, the fractional change in amplitude $\Delta A/A$ due to precessional
modulation is of order \cite{cord93}
\begin{equation}
\frac{\Delta A}{A} \approx \frac{\Delta \Theta}{W},
\end{equation}
where $\Delta \Theta$ denotes the change in polar angle of ${\bf
\hat{m}}$ over a free precession period.  We have
\begin{equation}
\label{deltaTheta}
\cos \Theta = {\bf \hat{m}}_{z} 
            = \sin \theta \sin \psi \sin \chi + \cos \theta \cos \chi,
\end{equation}
from which we find the (obvious) results that $\Theta$ has a maximum
value of $\chi + \theta$ and a minimum value of $|\chi - \theta|$.  It
follows that $\Delta \Theta$ is approximately equal to  $\theta$ or $\chi$,
whichever is smaller.

Pulsar duty cycles (i.e. angular widths of the beam) are typically of
order of $10^{\circ}$ \cite{lg98}, so we will parameterise according
to
\begin{equation}
\label{ampmod}
\frac{\Delta A}{A} \approx 6 \times 10^{-3} 
                   \left(\frac{\theta}{10^{-3}}\right)
                   \left(\frac{10^{\circ}}{W}\right),
\end{equation}
assuming $\theta < \chi$.  This modulation will occur at the body
frame free precession frequency.

\subsubsection{Polarisation modulation}
\label{sect:pm}

As above, we will assume a pulsar beam structure symmetric about ${\bf
\hat{m}}$.  In addition, we will take the polarisation model described in
Lyne \& Graham-Smith (1998, section 12.2).  In this model the polarisation
vector of the observed radiation is parallel to the magnetic field line
where the radiation was produced.  Then the time variation of the linear
polarisation angle, $\lambda$, can be calculated (see their equation 12.2).
It is this quantity that is measured by observers.  In the absence of free
precession this angle varies rapidly at the spin frequency.  Free
precession will cause an additional modulation at precession period $P_{\rm
fp}$.  A useful diagnostic for observers is the maximum rate of change of
this angle with time, $\dot{\lambda}_{\rm max}$, which occurs when the
dipole axis lies closest to the observers line-of-sight.  Let $i$ denote
the inclination angle, i.e. the angle between ${\bf J}$ and the
line-of-sight.  Using the equation of Lyne \& Graham-Smith for $\lambda$ we
then find (for $\theta < \chi$) a fractional modulation in
$\dot{\lambda}_{\rm max}$ of
\begin{equation}
\frac{\Delta \dot{\lambda}_{\rm max}}{\dot{\lambda}_{\rm max}} 
 = \frac{2 \sin i}{\sin \chi \sin (i - \chi)}  \theta.
\end{equation}

\subsection{Effect of free precession on the pulsar signal: 
            Electromagnetic torque effects}
\label{sect:efpps:et}

The calculations of section \ref{sect:efpps:ge} considered a precessing
symmetric top entirely free of torques.  However, the fact that the star is
visible as a pulsar means that it will be acted upon by an electromagnetic
torque, and this should be included in the calculation.  This torque will
not affect the amplitude and polarisation arguments.  However, provided
that the torque is a function of the spin rate and orientation of ${\bf
m}$, i.e. of $\dot{\Phi}$ and $\Theta$, the spin-down torque will be
modulated by the precession.  This modulation in the torque must be taken
into account when calculating the phase residuals.  We will call this
\emph{electromagnetic} modulation.  The affect on the phase residuals of
this varying torque has been considered analytically by Jones (1988) and
Cordes (1993) for general torque functions, and numerically by Melatos
(1999) for the \nocite{deut55} Deutsch (1955) torque.  We, however, will
give a simple description based on the vacuum point-dipole spin-down torque
so that
\begin{equation}
\label{vacdipole}
\ddot{\Phi} = k \dot{\Phi}^{3} \sin^{2} \Theta,
\end{equation}  
where $k$ is a negative constant.  The fractional change in spin-down rate
due to precession is given by
\begin{equation}
\label{fracmodPDd}
\frac{\Delta \ddot{\Phi}}{\ddot{\Phi}} \approx   3\frac{\Delta \dot{\Phi}}{\dot{\Phi}} 
                           + 2\frac{\Delta(\sin \Theta)}{\sin \Theta}.
\end{equation}
The prefix $\Delta$ denotes the departure of the respective quantity from
the smooth non-precessing power law spin-down.  We will consider the case
$\theta < \chi$.  From equation (\ref{DPdsmalltheta}) we see that the first
term is of order $\epsilon_{\rm eff} \theta$.  Using equation
(\ref{deltaTheta}) it is easy to show that
\begin{equation}
\sin^{2} \Theta \approx \sin^{2} \chi - 2\theta \sin \chi \cos \chi \sin \psi
\end{equation}
so that the second term of (\ref{fracmodPDd}) is of order $\theta$,
and therefore is the dominant one.  We then have
\begin{equation}
\Delta \ddot{\Phi} \approx -2k\Omega^{3} \theta \sin \chi \cos \chi \sin \psi.
\end{equation}
All the quantities on the right hand side are constant, apart from the
angle $\psi$, which is a linear function of time: $\psi = \dot \psi t$.
We can integrate once to get the perturbation in frequency
\begin{equation}
\Delta \dot{\Phi} \approx 2k\Omega^{3} \sin \chi \cos \chi \cos \psi \frac{\theta}{\dot{\psi}},
\end{equation}
and once more to obtain the phase residual
\begin{equation}
\Delta {\Phi} \approx 2k\Omega^{3} \sin \chi \cos \chi \sin \psi 
                      \frac{\theta}{{\dot{\psi}}^2}.
\end{equation}
Now use $|\dot{\psi}/\dot{\Phi}| = \epsilon_{\rm eff}$ to give
\begin{equation}
\Delta \Phi = \frac{1}{\pi} \cot \chi \frac{\theta}{\epsilon_{\rm eff}^{2}} 
      \frac{P}{\tau_{\rm e}},
\end{equation}
where $P$ denotes the spin period and $\tau_{\rm e} = |\ddot{\Phi}/
\dot{\Phi}|$ is the spin-down timescale.  For example, if we consider a
star where superfluid pinning is not operative, and where only the crust
participates in the free precession, we can put $\epsilon_{\rm eff} \approx
b \epsilon_{\Omega}I_{\rm star}/I_{\rm crust}$ (assuming $\epsilon_{0}
\approx \epsilon_{\rm fluid}$) to give
\begin{equation}
\label{DPtwo}
\Delta \Phi = 25  \cot \chi  \left(\frac{\theta}{10^{-3}}\right)
                  \left(\frac{10^{-5}}{b}\right)^{2}
                  \left(\frac{P}{30 \, \rm ms}\right)^{3}
\end{equation}
\[
\hspace{1cm}   \left(\frac{10^{3} \, \rm yrs}{\tau_{\rm e}}\right)
               \left(\frac{I_{\rm crust}/I_{\rm star}}{0.015}\right)^{2}.
\]
We have parameterised with a young pulsar in mind.  This should be compared
with the torque-free residual of equation (\ref{DPsmalltheta}).  We see that
for the parameterisation above the electromagnetic torque variation has
greatly amplified the residual.  Note, however, that for millisecond
pulsars which have $\tau_{\rm e} \la 10^{9}$ years the torque variations
are unimportant.

Note also that for the parameterisations above, as $\Delta \Phi$ is greater
than $2\pi$, this phase variation makes it difficult to detect such
pulsars---any search algorithm that integrates radio data assuming a
constant frequency source will go out of phase with the signal in an
interval of the order of the free precession period.  However, pulsar
physicists do take such a modulation into account when performing data
analysis when they search for binary pulsars, as a sinusoidal phase
variation is also produced by a binary orbit: Small-angle free precession
and nearly-circular binary orbits \emph{both} produce sinusoidal phase
residuals, to leading order in wobble angle and orbital eccentricity,
respectively.  As pointed out by Nelson et al.\ (1990), in order to
distinguish between the two models it is necessary to include higher order
terms.  Then the residuals have different forms, allowing differentiation
between the two models.  Also, free precession will almost certainly lead
to an amplitude modulation in phase with the timing residuals, whereas a
binary companion would only produce amplitude modulation if it happened to
cut the observer's line-of-sight onto the pulsar beam, providing another
means of differentiation.

We will present the phase residual in one more form.  If we put
$\dot{\psi} = 2\pi/P_{\rm fp}$ we obtain
\begin{equation}
\label{compare}
\Delta \Phi = \frac{1}{\pi} \cot \chi
                    \left(\frac{P_{\rm fp}}{P}\right)
                    \left(\frac{P_{\rm fp}}{\tau_{\rm e}}\right)
                    \theta.
\end{equation}

\section{Analysis of observations of free precession}
\label{sect:aooofp}

Having set out our free precession model in some detail, and described how
free precession modulates the electromagnetic signal of a pulsar, we will
now turn to the problem of extracting useful information from the pulsar
observations.  First we will assemble the necessary equations.

\subsection{Formulae required to extract source parameters from
observations} 

The problem divides into two parts: Extracting the wobble angle $\theta$,
and extracting information concerning the structure of the star, such as
its reference shape or the amount of pinned superfluid.

The extraction of the wobble angle is relatively straightforward.  We need
only invert the equations of the previous section.  Comparing the geometric
phase residual of equation (\ref{DPsmalltheta}) with the electromagnetic
torque residual of equation (\ref{compare}), we see that the
electromagnetic torque significantly amplifies the geometric residual when
\begin{equation}
\label{DPratio}
\frac{1}{\pi}  \left(\frac{P_{\rm fp}}{P}\right)
                    \left(\frac{P_{\rm fp}}{\tau_{\rm e}}\right) \gg 1.
\end{equation}
When this equality is satisfied equation (\ref{compare}) applies, and
the wobble angle can be extracted (up to a factor of $\tan \chi$) from
the observed values of spin period, free precession period and phase
residual magnitude according to
\begin{equation}
\label{thetaobs}
\theta = \pi \left(\frac{P}{P_{\rm fp}}\right)
             \left(\frac{\tau_{\rm e}}{P_{\rm fp}}\right) 
             \Delta \Phi \tan \chi.
\end{equation}
Note that if the spin-down torque is a steeper function of $\Theta$ than
the vacuum dipole model predicts (equation \ref{vacdipole}) the wobble
angle as estimated by the above equation will be an overestimate.  If the
torque is not as steep as assumed, then the calculated wobble angle will be
an underestimate.  

When the inequality is reversed the phase residuals are described
accurately by the geometric variation, and equation (\ref{DPsmalltheta})
applies.  Then the wobble angle can be extracted (up to a factor of $\tan
\chi$) from the value of $\Delta \Phi$ according to
\begin{equation}
\label{geothetaobs}
\theta = \Delta \Phi \tan \chi.
\end{equation}

Information concerning the structure of the star is more difficult to
obtain.  We hope to extract this information from equation (\ref{compeeff}),
where the quantity $\epsilon_{\rm eff}$ is simply the ratio of spin and
modulation periods, $P/P_{\rm fp}$.  Ideally, we would like to extract the
three quantities $\Delta I_{\rm d}$ (the deformation in the moment of
inertia tensor caused by Coulomb forces), $I_{\rm SF}$ (the moment of
inertia of the pinned superfluid), and $I_0$ (the part of the moment of
inertia which participates in the free precession).  Clearly, the problem is
underdetermined.  However, the quantity $\Delta I_{\rm d}/I_{\rm star}$ is,
for a given oblateness $\epsilon_0$, constrained by the equation of state
according to equation (\ref{DId}), so that:
\begin{equation}
\label{finaleff}
\epsilon_{\rm eff} =   \frac{P}{P_{\rm fp}}
                   = b \epsilon_0 \frac{I_{\rm star}}{I_0} 
                     + \frac{I_{\rm SF}}{I_0}.
\end{equation}  
The problem then becomes one of extracting $\epsilon_0$, $I_{\rm SF}$, and
$I_0$.  The quantity $\epsilon_0$ is, for a spinning-down star, almost
certainly greater than or equal to the actual oblateness $\epsilon_{\rm
fluid}$ (given by equation \ref{epsilonfluid}), and cannot exceed
$\epsilon_{\rm fluid}$ by more than $u_{\rm break}$.  Also, if our
understanding of crust-core coupling is correct, $I_0$ is simply the
crustal moment of inertia.  Given these assumptions, the above equation can
be used to place limits on $I_{\rm SF}$, the least certain of all the
stellar parameters.

However, we will employ a slightly simpler strategy, for the following
reason: If, as some glitch theories require, a few percent of the stars'
moment of inertia was made up of a superfluid pinned to the inner crust,
the effective oblateness parameter would be of order unity, so that $P_{\rm
fp} \sim P$.  If the observations below really do represent free
precession, this prediction has failed completely.  We will therefore
proceed as follows.  We will begin by setting $I_{\rm SF}$ to zero,
allowing us to extract a reference oblateness:
\begin{equation}
\label{npeoobs}
\epsilon_{0} = \epsilon_{\rm eff} \frac{1}{b} 
               \frac{I_0}{I_{\rm star}} 
               \hspace{1cm} ({\, I_{\rm SF}=0 \rm \, \, limit}). 
\end{equation}
In the case where $I_0 = I_{\rm crust}$ we can use equations
(\ref{bestimate}) and (\ref{Iratio}) to give:
\begin{equation}
\label{eoobs}
\epsilon_{0} = 10^3 \epsilon_{\rm eff} 
                    \frac{M_{1.4}}{R_6}
                    \hspace{1cm} ({\, I_{\rm SF}=0 \rm \, \, limit}).
\end{equation}
This can then be tested against inequality (\ref{ineqobs}).  There are
three possible cases.

Case I: If the inequality is violated because $\epsilon_0 < \epsilon_{\rm
fluid}$, our model requires modification.  The most likely modification
is that more than just the crust participates in the motion, so that $I_0 >
I_{\rm crust}$.  This would be a sign of stronger crust-core coupling than
anticipated.  Setting $I_{\rm SF}$ to a non-zero value would only serve to
increase $I_0$.  

Case II: If the inequality is satisfied, then the observation is consistent
with superfluid pinning not playing a role in free precession.  The
possibility that superfluid pinning is making up some part of the total
$\epsilon_{\rm eff}$ remains, but $I_{\rm SF}/I_0$ can be no larger than
$\sim \epsilon_{\rm eff}$.

Case III: The inequality is violated because $\epsilon_0 - \epsilon_{\rm
fluid} > u_{\rm break}$.  Of course, $u_{\rm break}$ is unknown, but is
surely less than $10^{-2}$ \cite{rude92}.  If the inequality is violated even for this
large breaking strain, then superfluidity is playing the dominant role in
determining the free precession frequency.  In this case $\epsilon_{\rm
eff} \sim I_{\rm SF}/ I_0$, or equivalently
\begin{equation}
\label{npSFobs}
\frac{I_{\rm SF}}{I_{\rm star}} 
       = \epsilon_{\rm eff} \frac{I_0}{I_{\rm star}}
         \hspace{1cm} ({\, \epsilon_0 =0 \rm \, \, limit}).
\end{equation}
Again parameterising with $I_0 = I_{\rm crust}$ using equation
(\ref{Iratio}) we obtain:
\begin{equation}
\label{SFobs}
\frac{I_{\rm SF}}{I_{\rm star}} = 1.5 \times 10^{-2} \epsilon_{\rm eff} 
           \frac{R_6^4}{M_{1.4}^2} 
          \hspace{1cm} ({\, \epsilon_0 =0 \rm \, \, limit}).
\end{equation}

\subsection{Analysis of observations}
\label{sect:aoo}

We will now apply our free precession model to the proposed observations of
free precession to have appeared in the literature.  The collection below
does not represent \emph{all} of the proposed candidates: We have not
included observations where only the (proposed) free precession timescale
$P_{\rm fp}$ has been measured, and not the (average) spin period $P$.
Also, we have not included sources where only one modulation cycle in the
pulsations (or less) has been observed.  In restricting our sample in this
way, we will confine our attention to sources where the evidence for a
modulation in the pulsations is reasonably secure, and where the
observational data is good enough for source parameters to be estimated.

\subsection*{PSR B0531+21 (T\lowercase{he} C\lowercase{rab pulsar})}

Lyne, Pritchard and Smith (1988) observed a variation in the phase residual
of the Crab pulsar of about 1.9 radians, with period 20 months.  Jones
(1988) suggested free precession as the cause, and pointed out that the
residuals would be dominated by the electromagnetic torque variation.
Equation (\ref{thetaobs}) then gives $\theta \approx 5 \times 10^{-6} \tan
\chi$ radians.  

Equation (\ref{eoobs}) gives a reference oblateness of $6 \times 10^{-7}$,
while equation (\ref{epsilonfluid}) gives $\epsilon_{\rm fluid} = 2 \times
10^{-4}$.  This violates the inequality (\ref{ineqobs}) (case I)---the star
would seem to have a reference shape much less oblate than its current
actual shape.  Even if the whole star is assumed to participate in the free
precession, so that $I_0 = I_{\rm star}$, we find $\epsilon_0 = 4.3 \times
10^{-5}$ (equation \ref{npeoobs}), so that inequality (\ref{ineqobs}) is
still violated.  Therefore, it is not possible to make this observation fit
our free precession model for any sensible stellar parameters.  Given that
the modulation has not been detected in subsequent observations, it seems
unlikely that free precession was detected.

More recently, evidence has been presented for a 60s modulation in the Crab
pulsar's amplitude and phase residual, both in the optical band
\cite{cg96a,cg96b,cgc97}.  This is by far the shortest free precession
period to have been proposed in the literature.  \v{C}ade\v{z} \&
Gali\v{c}i\v{c} report a phase residual amplitude of $\Delta \Phi = 1.2
\times 10^{-3}$ radians.  For $P_{\rm fp} = 60$s electromagnetic torque
amplification is insignificant.  Equation (\ref{geothetaobs}) gives $\theta
\approx 1.2 \times 10^{-3} \tan \chi$ radians.  They also report a fractional
amplitude modulation of $6 \times 10^{-3}$.  Equation (\ref{ampmod}) then
gives $\theta \approx 1.1 \times 10^{-4}$.  These two $\theta$ estimates
would agree for $\chi \sim 0.1$ radians.
 
In the absence of pinning, equation (\ref{eoobs}) gives a reference
oblateness of 0.55.  This is a large value, violating (\ref{ineqobs}) (case
III) even for $u_{\rm break}=10^{-2}$.  In other words, the reference
oblateness is too large to be accounted for by Coulomb deformation alone.
It is therefore necessary to invoke superfluid pinning: Using equation
(\ref{SFobs}), we see a fraction of order $10^{-5}$ of the star's moment of
inertia needs to be pinned to fit the data.  However, a subsequent search
\cite{getal00} has failed to confirm the modulation, weakening the
precession hypothesis.

\subsection*{PSR B0833-45 (The Vela pulsar)}

Deshpande \& McCulloch (1996) have presented evidence for a 165d variation
in the Vela's amplitude in the radio band.  The fractional modulation is of
order 1/2.  They have also investigated the difference in time-of-arrival
of the pulses at two different radio frequencies, and found a 330d
variation.  The most natural explanation of this latter variation, and
probably therefore of the former too, would be connected with refractive
scintillations \cite{lg98} due to the inter-stellar medium.  However, the
authors suggested free precession as the cause.  It is not easy to see how
precession could cause the variation in the time-of-arrival difference
between the radio frequencies.  We will therefore concentrate on the
amplitude variation.  Equation (\ref{ampmod}) gives $\theta \approx 0.1$
radians $\approx 6^{\circ}$.  

Assuming no superfluid pinning, the 165d modulation in the Vela gives a
reference oblateness of $\epsilon_{0} = 6 \times 10^{-6}$, a factor of 4
less than the calculated actual oblateness, $\epsilon_{\rm fluid} = 3
\times 10^{-5}$.  We therefore see that the inequality (\ref{ineqobs}) is
violated (case I)---the star is more nearly spherical than we would expect.
In order to increase $\epsilon_0$ to its minimum acceptable value of
$\epsilon_{\rm fluid}$, at least $6\%$ of the star would need to participate
in the free precession.  Inclusion of superfluid pinning increases this value.

\subsection*{PSR B1642-03}

Cordes (1993) has reported evidence for $10^{3}$d variations in the pulse
shape and timing residuals of pulsar B1642-03, both in the radio band.  Three
cycles have been observed (see figure VI of his paper), although the
profile is far from sinusoidal.  The fractional pulse shape modulation was
approximately 0.05, so that equation (\ref{ampmod}) leads to $\theta
\approx 9 \times 10^{-3}$ radians.  The phase residual amplitude is
difficult to identify as it seemed to increase over the observation period,
but a value $\Delta \Phi \approx 0.025$ is a reasonable average.  This
pulsar has residuals dominated by the electromagnetic torque variation
(equation \ref{DPratio}), so that equation (\ref{thetaobs}) gives $\theta
\approx 8 \times 10^{-4} \tan \chi$ radians.  These two estimates are
consistent for $\chi \approx 0.1$ radians.

For the reference oblateness in the zero pinning limit, equation
(\ref{eoobs}) gives $\epsilon_{0} = 5 \times 10^{-6}$, while equation
(\ref{epsilonfluid}) gives $\epsilon_{\rm fluid} = 10^{-6}$, so that
inequality (\ref{ineqobs}) is satisfied (case II).  Therefore, this
observation is consistent with zero superfluid pinning.  The calculated
reference oblateness is very close to the actual shape, suggesting that the
star is virtually unstrained.  This could imply a very low crustal breaking
strain of order $10^{-6}$.  The pinned superfluid component can make up no
more than $10^{-10}$ of the total stellar moment of inertia.

\subsection*{PSR B1828-11}

Very recently a 1009d periodicity has been reported in the radio
shape and residuals of pulsar B1828-11 \cite{sls00}.  Approximately four
cycles have been observed.  The pulse width is about $3^{\circ}$, and the
pulse shape changes drastically over one free precession period, suggesting
$\theta \sim 3^{\circ}$.

The phase residuals give, via equation (\ref{thetaobs}), a wobble angle of
only $3.5 \times 10^{-4}$ radians $= 0.02^\circ$, two orders of magnitude
less.  However, as is immediately apparent, there is a very strong 504d
periodicity in the data.  If we put $P_{\rm fp} = 504$d we obtain a $\theta
= 0.08^\circ$, still a factor of order 30 too small to agree with the
wobble angle derived from the amplitude modulation.  (Stairs et al. obtain
$0.3^\circ$ using a similar method of calculation, still one order of
magnitude smaller than the amplitude-derived value).  There is clearly a
problem in extracting the wobble angle of the source.

However, the fact that the data contains a strong periodicity at 504d
strongly suggests the following: The free precession period is indeed 1009
days, but the magnetic dipole lies very nearly orthogonal to the star's
deformation axis, i.e. $\chi \approx \pi/2$.  Then both the phase
modulation of equation (\ref{generalDP}) and the amplitude modulation
connected with equation (\ref{deltaTheta}) have significant components at
$2\dot \psi$.  To show this expand these equations to second order in
$\theta$. For the phase modulation:
\begin{equation}
\label{DPquad}
\Delta \Phi = - \theta \cot \chi \cos \psi
          - \frac{1}{4} \theta^2 \sin 2 \psi (1 + 2\cot^2 \chi),
\end{equation}
where the time dependence of the right hand side is due to $\psi= \dot \psi
t$.  The first term is the linear (in $\theta$) phase modulation at
$\dot{\psi}$, and the second the quadratic (in $\theta$) phase modulation
at $2 \dot \psi$.  The amplitude modulation for an unmagnetised star is due
to the variation in $\Theta$ given by
\[
\sin^2 \Theta =    \sin^2 \chi + \theta^2 \cos 2 \chi 
\]
\begin{equation}
\label{Thetaquad}
\hspace{2cm}      - 2 \sin \chi \cos \chi \sin \psi \theta
               -  \frac{1}{2} \theta^2 \sin^2 \chi \cos 2 \psi.
\end{equation}
The first two terms are uninteresting constants, the third the linear phase
modulation at $\dot{\psi}$, and the last the quadratic phase modulation at
$2 \dot \psi$.  The quadratic terms in (\ref{DPquad}) and (\ref{Thetaquad})
dominate the linear ones when:
\begin{equation}
\label{chibound}
\tan \chi > \frac{4}{\theta}.
\end{equation}
The calculation of section \ref{sect:efpps:et} for the phase modulation
for a magnetised star can then be repeated.  In the $\chi \rightarrow
\pi/2$ limit we obtain
\begin{equation}
\Delta \Phi = \frac{1}{4\pi} \frac{P_{\rm fp}^2}{\tau_{\rm e}P} 
              \theta^2 \cos 2 \psi
\end{equation}
We can then invert this relation to give the wobble angle.  Using $P_{\rm
fp}$ gives $\theta = 2^\circ$, in excellent agreement with the value
estimated using the amplitude modulation.  For consistency $\chi$ must then
satisfy (\ref{chibound}), which gives $\chi > 89^\circ$.  In other words,
for this scenario to apply, we require near perfect orthogonality between
the deformation axis and the magnetic dipole axis.

The spin and modulation periods of this star are, by coincidence, almost
identical to those reported above for PSR B1642-03, so the same conclusions
regarding the star's structure can be drawn, \emph{viz} that the
observation is consistent with our free precession model, with zero
superfluid pinning and a crustal strain of no more than $\sim 10^{-6}$.
The maximum amount of superfluid pinning allowed is of order $10^{-10}$.

\subsection*{Remnant of SN 1987A}
\label{sect:mobs}

Middleditch et al.\ (2000a,b) have recently presented evidence for a 2.14ms
optical pulsar in the remnant of SN 1987A.  The source was observed
intermittently between 1992 and 1997.  A modulation in the phase and
amplitude were detected, which the authors suggested may be due to free
precession.  The period of the modulation seemed to vary, spanning a range
of 935s to 1430s.  The amplitude of the phase modulation seemed to vary
from $\sim 48^\circ$ to $\sim 60^\circ$, although in some observing runs it
was possibly zero.  Middleditch et al.\ also suggest that the spin-down of
the pulsar is due to the gravitational radiation reaction associated with
the free precession.

We will examine the free precession and gravitational wave driven spin-down
using our model.  Crucially, the gravitational wave spin-down hypothesis
can be tested: The ratio of the spin to modulation periods will enable us
to calculate the deformation $\Delta I_{\rm d}$ of the star, while the
phase modulation will allow us to calculate the wobble angle $\theta$.
These can then be combined to give the gravitational wave spin-down, which
can then be compared to the observed value of the frequency derivative.

First we will extract a value for the wobble angle using the amplitude of
the phase residual, $\Delta \Phi$. This amplitude is highly variable,
implying a variable wobble angle.  At times the phase modulation seemed to
disappear.  If real, this disappearance would suggest that sometimes the
star simply spins about its symmetry axis without precessing.  When the
modulation was present, values in the range $\sim 48^\circ$ to $\sim
60^\circ$ were found.  Even if the spin-down were due entirely to a
magnetic dipole, inequality (\ref{DPratio}) shows that the phase residual
can be extracted while neglecting magnetic torque variations.  Then, if
$\Delta \Phi$ was small, equation (\ref{geothetaobs}) would give the wobble
angle (up to a factor of $\tan \chi$).  However, $\Delta \Phi$ is not
small, so the nonlinearised problem must be solved.  This is not possible
for a general value of $\chi$, but an example solution would be $\chi =
59^\circ$, $\theta = 80^{\circ}$ for $\Delta \Phi = 60^{\circ}$ (equation
\ref{generalDP}).  The key point is that $\theta$ is not small, and
the strains associated with such a large wobble angle would require a very
high crustal breaking strain.  Given that, to order of magnitude accuracy,
the precession-induced strain is of order $\theta \epsilon_{\rm fluid}$, we
have a strain of order $5 \times 10^{-2}$.  The breaking strain must be at
least as large as this, exceeding even the highest estimate of Ruderman
(1992), weakening the free precession hypothesis.

Now use equation (\ref{eoobs}) to extract the reference oblateness.  Using
a modulation period of 1009s we find $\epsilon_{0} = 2.1 \times
10^{-3}M/R$.  Using equation (\ref{epsilonfluid}) we find that the fluid
oblateness is $4.6 \times 10^{-2}R^3 /M$.  The inequality (\ref{ineqobs})
is violated (case I), even for high mass, small radius stars---the star is
more nearly spherical than we would expect if its crust solidified at a
rotation rate as high as 2.14ms.  This discrepancy is resolved only if a
significant portion of the star's total moment of inertia participates in
the free precession.  Using equation (\ref{npeoobs}) we see that we need
$I_0 = 0.33 I_{\rm star}$ for $\epsilon_0$ to be as large as $\epsilon_{\rm
fluid}$.  

Now we will test the gravitational wave spin-down hypothesis.  Balancing
the rate of loss of angular momentum of the star against the gravitational
flux we find \cite{cj00}:
\begin{equation}
\dot \Omega = \frac{32G}{5c^5} \Omega^5 
              \frac{(\Delta I_{\rm d})^2}{I_{\rm star}}
              \sin^2 \theta (\cos^2 \theta + 16 \sin^2 \theta)
\end{equation}
The quantity $\Delta I_{\rm d}$ can be extracted from observations: $\Delta
I_{\rm d}/I_{\rm 0} = P/P_{\rm fp}$, so that
\begin{equation}
\dot \Omega = \frac{32G}{5c^5} \Omega^5 
              \left(\frac{P}{P_{\rm fp}}\right)^2
              \frac{I_0^2}{I_{\rm star}}
              \sin^2 \theta (\cos^2 \theta + 16 \sin^2 \theta)
\end{equation}
Parameterising:
\[
\left(\frac{\dot f}{20 \, \mu Hz / \rm day}\right) 
  = 1.5 \sin^2 \theta (\cos^2 \theta + 16 \sin^2 \theta)
\]
\begin{equation}
    \hspace{2cm}
    \left(\frac{2.14}{P}\right)^3
    \left(\frac{10^3}{P_{\rm fp}}\right)^2
    \left(\frac{I_0/I_{\rm star}}{0.33}\right)^2
    M_{1.4} R_6^2.
\label{fdotwhole}
\end{equation}
For $\theta \sim 1$ and $I_0 = 0.33I_{\rm star}$ this is an order of
magnitude \emph{larger} than the observed value.  However, if we were to
disregard the $\theta \sim 1$ derived from the phase modulation, we could
use the above equation to calculate the wobble angle required for
gravitational waves to provide angular momentum balance.  We find $\theta =
0.44$ radians $= 25^\circ$.  This corresponds to a crustal strain of order
$\theta \epsilon_{\rm fluid} \sim 2 \times 10^{-2}$.  The breaking strain
must be at least as large as this.  This is an uncomfortably large value,
exceeding even the highest estimates \cite{rude92}.  If the whole star
participates in the free precession, so that $I_0 = I_{\rm star}$, the
wobble angle necessary to balance the spin-down is $9^{\circ}$,
corresponding to a slightly more plausible breaking strain of $8 \times
10^{-3}$.  Clearly, if the free precession and gravitational wave spin-down
hypotheses are correct, our model points towards most or all of the star
participating in the precession.

One of the main pieces of evidence Middleditch et al.\ cite to support the
gravitational wave spin-down hypothesis is that the frequency derivative
$\dot f$ correlates rather well with the inverse square of the long
modulation period $P_{\rm fp}$.  This is illustrated by figure 5 of
Middleditch et al.\ (2000b), where $\dot f$ varies by a factor of $\sim 4$,
while $P_{\rm fp}$ varies by a factor of 2.  As can be seen from equation
(\ref{fdotwhole}), this correlation is indeed predicted by this model,
\emph{but only for a fixed wobble angle}.  For the correlation to be
perfect, the wobble angle would have to remain exactly constant.  Referring
to their figure, we see that with the exception of a single data point, all
the error bars are consistent with the wobble angle remaining constant.
The maximum fractional variation in the angular function $\sin^2 \theta
(\cos^2 \theta + 16 \sin^2 \theta)$ consistent with $\dot f$ remaining
within the error bars is approximately 10\%, corresponding to a variation
away from $\theta = 9^\circ$ of less than $1^\circ$, for instance.  It is
not easy to see how the (unknown) process that leads to a factor of two
variation in the \emph{size} of the deformation $\Delta I_{\rm d}$ could
preserve the wobble angle to better than 10\%.

To sum up, the large phase residual of around $60^{\circ}$ cannot be
accounted for: In such a fast spinning star it would correspond to a
crustal breaking strain of around $5 \times 10^{-2}$, an implausibly large
value.  Setting this difficulty aside, in order to satisfy inequality
(\ref{ineqobs}) we found that at least one third of the star had to
participate in the free precession.  In order for this motion to provide
the necessary gravitational spin-down torque an even larger portion of the
star must precess.  If the whole star precesses, the gravitational wave
spin-down hypothesis requires a crustal breaking strain at least as large
as $8 \times 10^{-3}$.  If a smaller portion precesses, an even larger
breaking strain is required.

\subsection{Her X-1}

Her X-1 is a 1.24s X-ray pulsar in a 1.7d binary orbit about a
main-sequence star.  A third periodicity of 35d has been measured.  Some
authors have attributed this to forced precession of the accretion disk
\cite{pett75,gb76,bms90}, others to free precession of the star
\cite{brec72,spp98}, while others argue for a combination of both
\cite{tetal86,ketal00}.  If the 35d periodicity is due to free precession
the wobble angle must be large, of order unity \cite{spp98}.

Equation (\ref{eoobs}) gives a reference oblateness of $\epsilon_0 = 4.1
\times 10^{-4}$, while equation (\ref{epsilonfluid}) gives the star's
actual shape as $\epsilon_{\rm fluid} = 1.4 \times 10^{-7}$.  These values
are consistent with inequality (\ref{ineqobs}) (case II), so that the
observations are consistent with a crust-only precession with zero
superfluid pinning.  The pinned superfluid component can make up no more
than about $10^{-8}$ of the total moment of inertia.

\subsection{Summary and discussion of observations}

We will now comment upon our findings, which are summarised in table
\ref{obstable}.  
\begin{table*}
\begin{minipage}{150mm}
\caption{This table summarises stellar parameters calculated from the spin
and (proposed) free precession periods.  The quantity $\epsilon_{\rm eff}$
is simply the ratio of these, $P/P_{\rm fp}$.  The actual oblateness
$\epsilon_{\rm fluid}$ is calculated using equation (\ref{epsilonfluid}).
The quantity $\epsilon_0$ is the reference oblateness as calculated when
only the crust participates in the free precession, and there is no pinned
superfluid (equation \ref{eoobs}).  The quantity $I_{\rm SF}/I_{\rm star}$
is the fraction of the stellar moment of inertia made up of pinned
superfluid, assuming that only the crust participates in the free
precession, and that the reference oblateness is zero (equation
\ref{SFobs}).  Key to references: (1) Lyne et al.\ (1988); (2)
\v{C}ade\v{z} \& Gali\v{c}i\v{c} (1996a); (3) Deshpande \& McCulloch
(1996); (4) Cordes (1993); (5) Stairs, Shemar \& Lyne (2000); (6)
Middleditch et al.\ (2000a); (7) Trumper et al.\ (1986).}
\label{obstable}
\begin{tabular}{|l|l|l|l|l|l|} 
\hline

Object  & Reference & $\epsilon_{\rm eff}$ & $\epsilon_{\rm fluid}$ &
$\epsilon_0$  &  $I_{\rm SF}/I_{\rm star}$  \\ 

B0531+21 (Crab) & 1 & $6.4 \times 10^{-10}$ & 
$2 \times 10^{-4}$ & $6 \times 10^{-7}$ & $9.6 \times 10^{-12}$ \\  

B0531+21 (Crab) & 2 & $5.5 \times 10^{-4}$ &
$1.9 \times 10^{-4}$ & $5.5 \times 10^{-1}$ & $8.3 \times 10^{-6}$ \\

B0833-45 (Vela) & 3  & $6.2 \times 10^{-9}$ &
$2.7 \times 10^{-5}$ & $6.2 \times 10^{-6}$ & $ 9.3 \times 10^{-11}$ \\

B1642-03 & 4 & $4.5 \times 10^{-9}$ & 
$1.3 \times 10^{-6}$ & $4.5 \times 10^{-6}$ &  $6.8 \times 10^{-11}$ \\

B1828-11 & 5 & $4.7 \times 10^{-9}$ &  
$1.3 \times 10^{-6}$ & $4.7 \times 10^{-6}$ & $7.1 \times 10^{-11}$ \\

SN 1987A remnant & 6 & $2.1 \times 10^{-6}$ &   
$4.6 \times 10^{-2}$ & $2.1 \times 10^{-3}$ & $3.2 \times 10^{-8}$ \\

Her X-1 & 7& $4.1 \times 10^{-7}$  &  
$1.4 \times 10^{-7}$ & $4.1 \times 10^{-4}$ & $6.2 \times 10^{-9}$ \\  

\hline
\end{tabular}
\end{minipage}
\end{table*}
The key points are as follows:
\begin{itemize}
\item
Only one observation (Crab, Lyne et al.\ 1988) did not fit our free precession
model for any choice of parameters---the calculated reference shape was too
nearly spherical to have been formed by solidification earlier in the
Crab's life when the star spun more rapidly than it does today.
\item
Only one observation (Crab, \v{C}ade\v{z} \& Gali\v{c}i\v{c} 1996a)
required superfluid pinning for its explanation.  Assuming crust-only
precession, this pinned component must make up a fraction $10^{-5}$ of the
total stellar moment of inertia.
\item
The remaining five observations could be made to fit our model with the pinned
superfluid component set to zero, so that Coulomb forces provided the
deformation in the moment of inertia tensor.  Two observations required at
least part of the interior fluid to participate in the free precession.
Specifically, Vela required at least $6\%$ of the total stellar moment of
inertia to participate, while the SN 1987A remnant required at least
$33\%$.  The remaining observations were consistent with a crust-only free
precession.  The \emph{maximum} amount of pinned superfluid consistent with
the observations is typically $\sim 10^{-10}$ of the stars' total moment of
inertia.
\item
We were able to explain two otherwise puzzling features of the Stairs et
al.\ (2000) observation of PSR B1828-11 by assuming the dipole axis
to lie nearly orthogonally to the deformation axis ($\chi$ close to $\pi/2$).
Specifically, we were able to account for the presence of a strong 504d
periodicity in the 1009d modulation, and the apparent discrepancy between
the wobble angle as derived using the amplitude modulation and using the
phase residuals.
\end{itemize}

As stated, two observations require more than just the crust to follow the
free precession.  However, we wish to go further, and ask the following:
Are the observations consistent with part (or all) of the core liquid in
\emph{all} the neutron stars following the free precession?  Taking the
extreme case of total crust-core coupling, we set $I_0$ equal to the total
stellar moment of inertia in equations (\ref{npeoobs}) and (\ref{npSFobs}).
This simply serves to increase the values of $\epsilon_0$ and $I_{\rm
SF}/I_{\rm star}$ in the table by a factor of $I_{\rm star}/I_{\rm crust}
$, i.e. by a factor of $\approx 67 M_{1.4}^2 / R_6^4$ (equation
\ref{Iratio}).

In this case the qualitative conclusions that can be drawn are not very
different from before.  The Lyne et al.\ observation of the Crab is still
inconsistent with the free precession model (although the reference
oblateness is only a factor of 5 smaller than the fluid oblateness).  The
\v{C}ade\v{z} \& Gali\v{c}i\v{c} observation of the Crab still requires a
pinned superfluid, now making up a fraction $10^{-3}$ of the total moment
on inertia.  As stated previously, the SN 1987A observation can \emph{only}
be explained assuming a large fluid component in the free precession.  All
the remaining observations continue to fit our model---the only change is
that the maximum pinned superfluid component allowed is increased to values
typically of order $10^{-8}$, still much smaller than the $10^{-2}$
predicted by theory.

Another important question remains to be answered: Why are these stars
precessing in the first place?  What excited this motion?  The two main
candidates for isolated stars are glitches and electromagnetic torques.  A
glitch could excite free precession by occurring in a non-axisymmetric way,
suddenly shifting the principal axis of the moment of inertia tensor while
preserving the angular momentum \cite{lfe98}.  Alternatively, the spin-down
electromagnetic torque can amplify the wobble angle of an already
precessing star \cite{gold70}.  There may exist also an `anomalous'
electromagnetic torque \cite{jone88}, which does not contribute to the
spin-down, but will cause an initially non-precessing star to precess.
Both glitches and electromagnetic torques are important for young stars,
consistent with the rather young ages of the precession candidates. (All
the isolated stars in table \ref{obstable} have spin-down ages $P/2\dot P
\la 10^5$ years, with the exception of B1642-03, which has a spin-down age
of $3 \times 10^6$ years).  In the case of Her X-1, the accretion torque is
the obvious source of wobble excitation \cite{llps75}.

If electromagnetic torques are responsible, then we would expect almost all
sufficiently young pulsars to display signs of free precession.  In
contrast, if glitching is responsible, we would expect to observe free
precession in those stars which happened to have glitched recently, so that
the precession has not yet been damped away.  Only a fraction of
the young pulsars show signs of free precession, favouring the glitch
hypothesis.

\section{Conclusions}
\label{sect:conc}

As stated at the outset, our goal was to extract as much information as
possible from the handful of pulsar candidates for free precession.  To
this end we built a free precession model capturing (we hope) the main
features of the problem.  Our model contained two coupling mechanisms
between the crust and neutron fluid core.  One was \emph{inertial
coupling}, where the fluid core effectively `pushed' on the surrounding
crust.  The other was a frictional coupling due to the scattering of
electrons off neutron vortices.  We argued that even when these effects are
taken into account, the neutron fluid core is not expected to follow the
precession of the crust.

When we compared our model against the observations we found that the
wobble angle of the candidates was typically small, less than a degree or
so.  Of greater interest was the information that might be extracted from
the ratio of the spin and free precession periods, $\epsilon_{\rm eff}$.
This ratio depends on the details of the star's structure, specifically on
the geological history of the crust (parameterised by $\epsilon_0$), on the
amount of pinned superfluid ($I_{\rm SF}$), and on the portion of the star
that participates in the free precession ($I_0$).  It was not possible to
examine the three parameters independently, as they \emph{all} contribute
to the observed $\epsilon_{\rm eff}$ (see equation \ref{compeeff}).

However, by assuming reasonable values for crustal strength ($b$) and
breaking strain ($u_{\rm break}$), a few general conclusions could be
drawn.  Firstly, if only the crust participates in the free precession,
superfluid pinning need be invoked to explain only one of the observations.
The others were consistent with no pinning at all, with an upper bound on
$I_{\rm SF}$ typically of $10^{-10}$ of the total stellar moment of
inertia.  In the (in our opinion less likely) case where the whole star
precesses, the maximum pinned component was found to be typically $10^{-8}$
of $I_{\rm star}$.  Both these values are many orders of magnitude smaller
than predicted by some glitch theories.  Clearly, if the observations
considered here really do represent free precession, superfluid pinning
theory, at least as it affects free precession, is in radical need of
reworking.

\section*{ACKNOWLEDGEMENTS}

It is a pleasure to thank Curt Cutler and Bernard Schutz for stimulating
discussions during the course of this work.  This work was supported by
PPARC grant PPA/G/1998/00606.


\begin{thebibliography}{99}

\bibitem[\protect\citename{Alpar \& Sauls }1988]{as88}
Alpar A., Sauls J. A., 1988, {\em Ap. J.} {\bf 327} 723

\bibitem[\protect\citename{Bisnovatyi-Kogan \& Kahabka }1993]{bk93}
Bisnovatyi-Kogan G. S., Kahabka P., 1993, 
{\em Astron. Astrophys. } {\bf 267} L43

\bibitem[\protect\citename{Bisnovatyi-Kogan et al.\ }1990]{bms90}
Bisnovatyi-Kogan G. S., Mersov G. A., ShefferE. K., 1990, 
{\em Sov. Astron.} {\bf 34(1)} 44

\bibitem[\protect\citename{Bondi \& Gold  }1955]{bg55}
Bondi H., Gold T., 1955, {\em MNRAS} {\bf 115} 41

\bibitem
[\protect\citename{Brecher }1972]
{brec72}
Brecher K., 1972, {\em Nature} {\bf 239} 325

\bibitem
[\protect\citename{\v{C}ade\v{z} \& Gali\v{c}i\v{c} }1996a]
{cg96a} 
\v{C}ade\v{z} A., Gali\v{c}i\v{c} M., 1996, {\em Astron. \& Astrophys. }
{\bf 306 } 443

\bibitem
[\protect\citename{\v{C}ade\v{z} \& Gali\v{c}i\v{c} }1996b]
{cg96b} 
\v{C}ade\v{z} A., Gali\v{c}i\v{c} M., 1996, in Johnston S., Walker M. A.,
Bailes M., eds, ASP Conference Series, Vol 105, American Society of the
Pacific

\bibitem
[\protect\citename{\v{C}ade\v{z} \& Gali\v{c}i\v{c} \& Calvani }1997]
{cgc97} 
\v{C}ade\v{z} A., Gali\v{c}i\v{c} M., Calvini M., 1997, {\em Astron. \&
Astrophys. } {\bf 324 } 1005

\bibitem
[\protect\citename{Cordes }1993]
{cord93}
Cordes J. A., 1993, in Phillips J. A. and Thorsett J. E., eds ASP
Conference Series, Vol 36

\bibitem[\protect\citename{Cutler \& Jones }2000]{cj00}
Cutler C., Jones D. I., 2000 To appear in {\em Phys. Rev. D}

\bibitem
[\protect\citename{Deshpande \& McCulloch  }1996]
{dm96} 
Deshpande A. A., McCulloch P. M., 1996, in Johnston S., Walker M. A.,
Bailes M., eds, ASP Conference Series, Vol 105, American Society of the
Pacific, p. 101

\bibitem[\protect\citename{Deutsch }1955]{deut55}
Deutsch A. J., 1955, {\em Annales d'Astrophysique} {\bf 18(1)} 1

\bibitem
[\protect\citename{Easson }1979]
{eass79}
Easson I., 1979, {\em Ap. J.} {\bf 228} 252

\bibitem
[\protect\citename{Gerend \& Boynton }1976]
{gb76}
Gerend D. \& Boynton P. E., 1976, {\em Ap. J.} {\bf 209} 562

\bibitem[\protect\citename{Gil et al.\ }1993]{gjk93}
Gil J. A., Jessner A., \& Kramer M., 1993 
{\em Astron. \& Astrophys.} {\bf 271} L17

\bibitem[\protect\citename{Glendenning }1995]{glen95}
Glendenning N., 1995, {\em Ap. J.} {\bf 440} 881

\bibitem[\protect\citename{Glendenning }1997]{glen97}
Glendenning N., 1997, Compact Stars. Springer

\bibitem[\protect\citename{Golden et al.\ }2000]{getal00}
Golden A. et al.\ 2000, To appear in {\em Astron. \& Astrophys.} 

\bibitem
[\protect\citename{Goldreich }1970]
{gold70}
Goldreich P., 1970, {\em Ap. J.} {\bf 160} L11

\bibitem
[\protect\citename{Jones }2000]
{jone00}
Jones D. I., 2000, PhD. Thesis, University of Wales, Cardiff

\bibitem
[\protect\citename{Jones }1988]
{jone88}
Jones P. B., 1988, {\em MNRAS} {\bf 235} 545

\bibitem
[\protect\citename{Ketsaris et al.\ }2000]
{ketal00}
Ketsaris et al.\, 2000, astro-ph/0010035

\bibitem
[\protect\citename{Lamb et al.\ }1975] 
{llps75} 
Lamb D. Q., Lamb F. K., Pines D., Shaham J., 1975, 
{\em Ap. J.} {\bf 198} L21

\bibitem[\protect\citename{Lamb }1952]{lamb52}
Lamb H., 1952, Hydrodynamics (Sixth Edition). Cambridge University Press

\bibitem[\protect\citename{Landau \& Lifshitz }1976]{ll76}
Landau L. D., Lifshitz E. M., 1976, Mechanics, 3rd Edition.
Butterworth-Heinemann Ltd.

\bibitem
[\protect\citename{Link et al.\ }1998]
{lfe98}
Link B., Franco L. M., Epstein R. I., 1998 {\em Ap. J.} {\bf 508} 838

\bibitem
[\protect\citename{Lyne \& Graham-Smith }1998]
{lg98}
Lyne A. G., Graham-Smith F., 1998, Pulsar Astronomy.  Cambridge University
Press

\bibitem
[\protect\citename{Lyne et al.\ }1988]
{lps88}
Lyne A. G., Pritchard R. S., Smith F. G., 1988, {\em MNRAS} {\bf 233} 667

\bibitem[\protect\citename{Melatos }1999]{mela99}
Melatos A., 1999, {\em Ap. J.} {\bf 519} L77

\bibitem
[\protect\citename{Middleditch et al.\ }2000a]
{metal00a}
Middleditch J., et al.\, 2000, {\em New Astronomy} {\bf 5} no.\ 5, 243

\bibitem
[\protect\citename{Middleditch et al.\ }2000b]
{metal00b}
Middleditch J., et al.\, 2000, astro-ph/0010044

\bibitem
[\protect\citename{Munk \& McDonald }1960]
{mm60}
Munk W. H., McDonald G. J. F., 1960, The Rotation of the Earth. 
Cambridge University Press

\bibitem[\protect\citename{Nelson, Finn \& Wasserman }1990]{nfw90}
Nelson R. W., Finn L. S., \& Wasserman I., 1990, {\em Ap. J.} {\bf 348} 226

\bibitem
[\protect\citename{Petterson }1975]
{pett75}
Petterson J. A., 1975, {\em Ap. J.} {\bf 210} L61

\bibitem
[\protect\citename{Pines \& Shaham }1972]
{ps72}
Pines D., Shaham J., 1972, {\em Phys. Earth and Planet. Interiors} 
{\bf 6} 103

\bibitem[\protect\citename{Ravenhall \& Pethick }1994]{rp94}
Ravenhall D. G., Pethick C. J., 1994 {\em Ap. J.} {\bf 424} 846

\bibitem[\protect\citename{Ruderman }1970]{rude70}
Ruderman M., 1970, {\em Nature} {\bf 225} 838

\bibitem
[\protect\citename{Ruderman }1992]
{rude92}
Ruderman M., 1992,
In {Structure and Evolution of Neutron Stars} p.353

\bibitem[\protect\citename{Shaham }1977]{shah77}
Shaham J., 1977, {\em Ap. J.} {\bf 214} 251

\bibitem
[\protect\citename{Shakura et al.\ }1998]
{spp98}
Shakura N. I., Postnov K. A. \& Prokhorov M. E., 1998, {\em Astron. \&
Astrophys. } {\bf 331} L37

\bibitem
[\protect\citename{Stairs et al.\ }2000]
{sls00}
Stairs I. H., Lyne A. G., Shemar S. L., 2000, {\em Nature} {\bf 406} 484

\bibitem
[\protect\citename{Trumper et al.\ }1986]
{tetal86}
Trumper J., Kahabka P., Ogelman H., Pietsch W. \& Voges W., 1986,
{\em Ap. J.} {\bf 487} L63

\bibitem[\protect\citename{Ushomirsky, Cutler \& Bildsten }2000]{ucb2000}
Ushomirsky G., Cutler C., Bildsten L., 2000 astro-ph/0001136


\end{thebibliography}
\end{document}